\documentclass[twocolumn,amsmath,amssymb,PRL]{revtex4}
\usepackage{color}
\usepackage{graphicx}
\usepackage{amsfonts}
\usepackage{amssymb}

\input epsf
\begin{document}
\title {Stability Spectroscopy of Rotons in a Dipolar Bose Gas}
\author{John P. Corson$^1$}
\author{Ryan M. Wilson$^2$}
\author{John L. Bohn$^1$}
\affiliation{$^1$JILA, NIST and Department of Physics, University of Colorado, Boulder, Colorado 80309-0440, USA}
\affiliation{$^2$JQI, NIST and Department of Physics, University of Maryland, Gaithersburg, Maryland 20899-8410, USA}
\date{\today}

\begin{abstract}
We study the stability of a quasi-one-dimensional dipolar Bose-Einstein condensate (dBEC) that is perturbed by a weak lattice potential along its axis. Our numerical simulations demonstrate that systems exhibiting a roton-maxon structure destabilize readily when the lattice wavelength equals either half the roton wavelength or a low roton subharmonic. We apply perturbation theory to the Gross-Pitaevskii and Bogoliubov de Gennes equations to illustrate the mechanisms behind the instability threshhold. The features of our stability diagram may be used as a direct measurement of the roton wavelength for quasi-one-dimensional geometries.
\end{abstract}
\maketitle

It is widely believed that ultracold, gaseous samples of bosonic atoms or molecules possessing sufficiently large dipole moments will exhibit internal structure reminiscent of the roton in superfluid helium \cite{Santos2003}. The basic phenomenology of the roton, in analogy with helium, is a {\it local minimum} in the quasiparticle dispersion $\omega(k)$. The existence of such a minimum is predicted to lead to a host of attendant phenomena in these dilute gases, including structured ground-state density profiles \cite{Ronen2007,Dutta2007,Lu2010}, reduced and anisotropic critical superfluid velocity \cite{Santos2003,Wilson2010,Ticknor2011}, enhanced sensitivity to external perturbations \cite{Wilson2008}, abrupt transitions in Faraday patterns \cite{Santos2010,Santos2012}, short-wavelength immiscibility phases \cite{Wilson2012}, and strongly-oscillatory two-body correlations on the roton length scale \cite{Sykes2012}. Signatures of the roton in Bragg spectroscopy of trapped dipolar Bose-Einstein condensates (dBECs) have been calculated in Ref. \cite{Blakie2012}. While these many exciting predictions are in principle observable in current experiments with highly magnetic atoms \cite{Griesmaier2005,Lu2011,Aikawa2012}, or in future experiments with electrically polar molecules \cite{Ospelkaus2008,Deiglmayr2008,Debatin2011}, rotons remain to be seen directly in dBECs \cite{Mottl2012}.

In a dipolar condensate, the roton represents a mode of finite wavelength that has an anomolously low energy -- hence the minimum in $\omega(k)$.  The location of this minimum is given by a momentum  approximately equal to $\hbar / l_{t}$, 
where $l_{t}=\sqrt{\hbar / m \omega_t}$ denotes the harmonic oscillator 
length of the tightest confinement of the dBEC
(usually along the polarization axis).  Without
this confinement, a homogeneous dBEC would be energetically unstable to collapse due to the
attraction between dipoles that are aligned head-to-tail. In the
presence of this confinement, the collapse is prevented by the zero-point
energy introduced by the confinement, at least up until a critical
dipole moment or density.  When this critical parameter is exceeded, the condensate collapses in localized
``clumps'' of size $\lambda_{\mathrm{rot}} \sim l_{t}$, that is, 
via a dynamical instability into the roton mode \cite{Bohn2009}.  
For dBECs that are just barely stable, the energy minimum
$\omega (k_{\mathrm{rot}})$ is achieved at the roton wave vector 
$k_{\mathrm{rot}}= 2 \pi / \lambda_{\mathrm{rot}}$.

A low-energy roton is therefore a mode that is linked intrinsically to condensate instability.  This link
suggests that the dBEC may respond nontrivially as
an object of conventional spectroscopy, that is, that it would absorb
and distribute energy differently from different wavelengths of an 
applied probe.  For example, if one imposes on 
a dBEC a weak potential of periodicity $\lambda_{\mathrm{L}}$ (via a 1D 
lattice beam \cite{Grimm2000}, for example), then one expects to trigger an instability 
most easily when $\lambda_{\mathrm{L}} \approx \lambda_{\mathrm{rot}}$.  In this Letter
we verify this conjecture via mean-field simulations of a quasi-1D (q1D)
dBEC.  Strikingly, we find structure in addition to this main peak, analogous
to nonlinear spectroscopy of atoms or crystals in strong laser fields.  
Namely, we see
hastened destabilization by probes with wavelengths that are integer multiples of $\lambda_{\mathrm{rot}}$, reminiscent of ``multiphoton'' scattering. We also observe a similar feature at the shorter wavelength $\lambda_{\mathrm{L}}=\lambda_{\mathrm{rot}}/2$, reminiscent of resonant Raman coupling. Since the stability of a condensate is easy to assess experimentally, the observation of such structures constitutes a direct {\it stability}-spectroscopic measurement of the roton wavelength as well as a novel signature of roton physics.

Consider a dBEC that is tightly confined in the $\hat y$ and $\hat z$ 
directions by a harmonic trap of frequency $\omega_t$, with no trapping 
potential in the $\hat x$ direction. The dipole moments of the constituent
atoms or molecules are polarized along $\hat z$. To this initially-stable
system is applied a probe in the form of an optical lattice potential
\begin{equation} \label{eq: pert}
U(x)/\hbar\omega_t = s \cos^2\left(\frac{k_{\mathrm{L}}}{2}x\right)
\end{equation}
whose periodicity $\lambda_{\mathrm{L}} = 2\pi/k_{\mathrm{L}}$ and (dimensionless) lattice depth 
$s\equiv -\mathrm{Re}\{\alpha(\omega)\}I_0/2\epsilon_0c\hbar\omega_t \ge 0$ 
are tunable parameters. Such a potential can be realized using retroreflected 
or crossed off-resonant laser beams of peak intensity $I_0$ \cite{Grimm2000}. We take the single-mode approximation, assuming that the order parameter is a Gaussian of width $\ell_t=\sqrt{\hbar/m\omega_t}$ in the directions of tight confinement \cite{Petrov2000}. We suppose that there are $N=2L n_{1D}$ atoms spread over the periodic domain $x\in[-L,L]$, where $n_{1D}$ is the one-dimensional integrated density. Expressed in natural length ($l_t$) and energy ($\hbar \omega_t$) units, 
the Gross-Pitaevskii equation for this situation reads
\begin{equation} \label{eq: GP}
\begin{aligned}
\mu\psi(x) =& -\frac{1}{2}\partial_x^2\psi(x) + U(x) \psi(x) \\ & + N\int dx'\psi^*(x')\psi(x')V(x-x')\psi(x).
\end{aligned}
\end{equation}
The momentum-space form of the q1D interaction potential is
\begin{equation} \label{eq: VintK}
\begin{aligned}
V(k)=& 2 a_s +4 a_{dd}\Bigg[1-\\&\frac{3}{\sqrt{2}}\int\limits_0^{\infty}dw \mathrm{e}^{-w^2/2}\mathrm{g}\left(\sqrt{\frac{w^2+k^2}{2}} \right)  \Bigg]
\end{aligned}
\end{equation}
where $\mathrm{g}(q)\equiv q\exp(q^2)\mathrm{erfc}(q)$. Contact 
interactions contribute to $V$ via the scattering length $a_s$, and 
dipole-dipole interactions contribute via the dipole length 
$a_{dd}=md^2/3\hbar^2$, where $d$ is the dipole moment of the bosons. 
We remark that $V(k)$ decreases monotonically as a function of $k$. 
The q1D chemical potential $\mu$ is related to the three-dimensional 
chemical potential $\mu_{3D}$ via $\mu = \mu_{3D}-1$. 
In the absence of the probing potential $U(x)$, the chemical potential 
and order parameter are easily found to be $\mu^{(0)}=n_{1D}V(0)$ and 
$\psi^{(0)}=1/\sqrt{2L}$, respectively.

\begin{figure}
\includegraphics[width=0.9\columnwidth]{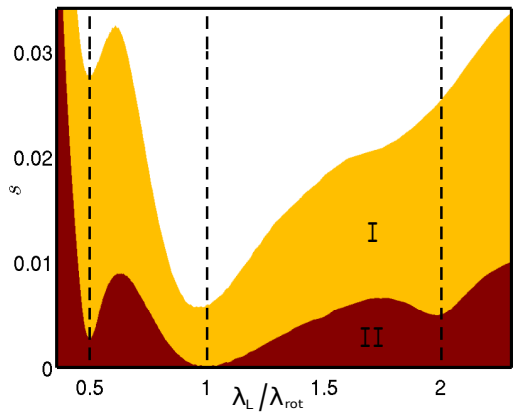}
\caption{Stability spectrum for a q1D dBEC in a weak lattice. The experimental parameters are for $^{164}$Dy trapped at $\omega_t = 2\pi \times 5$ kHz with density $n_{3D}\approx 10^{15}$ cm$^{-3}$ and dipole moment $d=10\mu_B$. In Case I (II), the scattering length is tuned to $a_s=-22a_0$ ($a_s=-24.5a_0$), yielding a roton wavelength of $900$ nm ($850$ nm). The vertical dotted lines denote $\lambda_{\mathrm{rot}}/2$, $\lambda_{\mathrm{rot}}$, and $2\lambda_{\mathrm{rot}}$, which identify the spectral features.}
\label{fig: stability}
\end{figure}

\begin{figure}
\includegraphics[width=0.9\columnwidth]{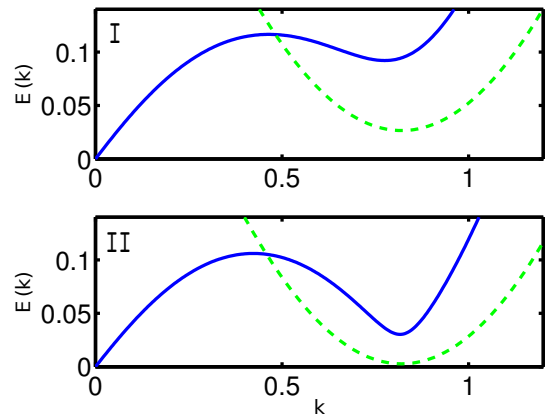}
\caption{(Color online) The green (dotted) and blue (solid) curves are respectively the Hartree-Fock and Bogoliubov spectra for the particular cases considered in Fig. \ref{fig: stability}.}
\label{fig: spectra}
\end{figure}

To determine the stability of the dBEC for a given lattice periodicity $\lambda_{\mathrm{L}}$ and intensity $s$, we solve Eq. \eqref{eq: GP}, and determine the excitation spectrum within the Bogoliubov approximation. Dynamical instability is indicated by a complex frequency in a given mode, which also identifies the mode that triggers the instability.  Numerical results are plotted in Fig. \ref{fig: stability}, which shows 
two sample stability spectra for dBECs consisting of $^{164}$Dy atoms.
Plotted is the intensity $s$ versus periodicity $\lambda_{\mathrm{L}}$ of the probe.
The shaded region below each curve represents the stable region, since
higher intensitites perturb the condensate more strongly and lead to
collapse.  In one case (I, yellow region) the scattering length is
assumed to be $a_s=-22a_0$, while in the other (II, brown), it is $a_s=-24.5a_0$.  These differences correspond to different roton spectra, as shown by the blue lines in Fig. \ref{fig: spectra}.  Case II has a ``softer'' roton than case I, i.e., a lower excitation energy $\omega(k_{\mathrm{rot}})$. Along the lines of Ref. \cite{Boudjemaa2012}, we have checked that our quantum and thermal depletions for both case I and case II, at an assumed temperature of $50$ nK, remain small compared to the condensate fraction. Similar results can be found for all other species of strongly dipolar condensates, although there is a tradeoff between density $n_{1D}$ and dipole length $a_{dd}$ for a given system to rotonize.

In both cases, the spectra clearly show three features, which
are more prominent for softer rotons: i) 
A {\it weak} laser drives the condensate to instability for probe periodicities
near the roton wavelength, accounting for the lowest dip in Fig. 
\ref{fig: stability}.  Indeed, the weakest laser required to
de-stabilize the condensate has periodicity $\lambda_{\mathrm{L}} = \lambda_{\mathrm{rot}}$. We emphasize that at the threshhold of stability, our simulations show that the lattice probe is not chopping the condensate into smaller, unstable pieces, but rather is weakly perturbing it. ii) A secondary dip in stability appears 
when the lattice spacing is twice the roton wavelength.  In terms of the
analogy to optical absorption in a nonlinear medium, this situation is reminiscent of the excitation of the resonant state by absorption of
two photons of half the energy required to excite the state directly.
iii) Finally, a third dip in stability is observed at half the roton 
wavelength. We have also observed these structures in full $3$D simulations of horizontal, cigar-shaped dBECs where the roton ``wavelength" is less well defined.

The origin of these features becomes clear within perturbation theory.
Rigorous perturbation theory for the Gross-Pitaevskii equation was 
worked out in Refs. \cite{Wilson2008,Taylor2003,Gaul2011,Liang2008}. Let
$\psi^{(0)}$ and $\mu^{(0)}$ represent the order parameter and chemical
potential in the absence of the probe $U(x)$.  Expanding
$\psi = \psi^{(0)} + \psi^{(1)}$, $\mu = \mu^{(0)} + \mu^{(1)}$,
one derives the first-order perturbation equation
\begin{equation} \label{eq: GP pert}
\mu^{(1)}\psi^{(0)}=-\frac{1}{2}\partial_x^2\psi^{(1)}+U\psi^{(0)}+2 n_{1D}\int dx' \psi^{(1)}V(x-x') .
\end{equation}
Given the condition $\langle \psi^{(1)}|\psi^{(0)}\rangle=0$, 
the solutions to \eqref{eq: GP pert} are easily found to be 
$\mu^{(1)}=s/2$ and
\begin{equation} \label{eq: GP1}
\psi^{(1)}=-\left(\frac{s}{2\varepsilon (k_{\mathrm{L}})}  \right)\frac{1}{\sqrt{2L}}\cos(k_{\mathrm{L}}x) .
\end{equation}
We have introduced the Hartree-Fock energy \cite{Pitaevskii2003} 
$\varepsilon(k)~=~k^2/2+2n_{1D}V(k)$.
Figure \ref{fig: spectra} compares $\varepsilon(k)$ to  the Bogoliubov 
spectra $\omega(k)~=~\sqrt{k^2/2\left(k^2/2+2n_{1D}V(k)\right)}$ 
for the two cases of dBEC considered in Fig. \ref{fig: stability}.

The Hartree-Fock spectrum of a soft-roton dBEC posesses a shallow minimum 
at $k\approx k_{\mathrm{rot}}$, and this minimal energy approaches zero as the 
roton softens. 
The effect of this low-lying mode is to strongly perturb the ground
state, as suggested by (\ref{eq: GP1}).  Indeed, a modest perturbation $\propto s$
becomes amplified in the density by a factor $s/\varepsilon(k_{\mathrm{rot}})$.
The $m^{\mathrm{th}}$ order of perturbation theory additionally introduces density modulations of the general form
$n_{1D}^{(m)}(x) \sim \frac{s^m}{\varepsilon(mk_{\mathrm{L}})}\cos(mk_{\mathrm{L}}x)$. 
That is, the roton can
be driven by overtones of the fundamental wave number of the lattice,
when $m k_{\mathrm{L}} \approx k_{\mathrm{rot}}$, or $\lambda_{\mathrm{L}} \approx m \lambda_{\mathrm{rot}}$.
The stability minimum corresponding to $m=2$ can be seen in Fig. 
\ref{fig: stability}.
Even away from one of these resonant conditions, the probe laser introduces
density modulations that manifest themselves in a 
mean-field potential $U_{\mathrm{mf}}(x)=N\int dx'|\psi|^2V(x-x')$  
which combines with the probe field itself to make a combined potential
\begin{equation} \label{eq: Uc}
\begin{aligned}
U_c(x) & =U(x)-\mu+U_{\mathrm{mf}}(x) \\& = \frac{s}{4}\frac{k_{\mathrm{L}}^2}{\varepsilon(k_{\mathrm{L}})}\cos(k_{\mathrm{L}}x)+\mathcal{O}(s^2) .
\end{aligned}
\end{equation}
Thus again, whatever influence the periodic probe potential has on propagation
of the excited states in the condensate, this effect is amplified for
probes in the vicinity of the roton (or a subharmonic thereof, as a higher-order effect).

There remains the task of explaining the feature in the stability
spectrum at $\lambda_{\mathrm{L}} \approx \lambda_{\mathrm{rot}}/2$.  This cannot be done by considering the ground state only.  Rather, it is necessary to explore how the spectrum of excited states is modified by the combined potential
energy $U_c$ (\ref{eq: Uc}).
The Bogoliubov de Gennes (BdG) equations describing the excitations can be
written compactly as
\begin{equation} \label{eq: BdG}
\begin{pmatrix}
H_0-\mu+C+X & X \\ -X & -H_0+\mu-C-X
\end{pmatrix}
\begin{pmatrix}
u \\ v
\end{pmatrix} = E
\begin{pmatrix}
u \\ v
\end{pmatrix}
\end{equation}
where $H_0 = -\frac{1}{2}\partial_x^2+U(x)$ is the free-particle Hamiltonian, 
$C[\chi](x) = U_{\mathrm{mf}}(x)\chi(x)$ describes direct interactions with the mean-field, and 
$X[\chi](x)=N\int dx'\chi(x')\psi(x')V(x-x')\psi(x)$ is an integral operator describing exchange interactions. The functions $u_j(x)$ and $v_j(x)$ define the usual 
Bogoliubov transformation of the quantum fluctuation field operator 
$\delta\hat\Psi =~\sum_j\left[\hat\alpha_ju_j+\hat\alpha^{\dagger}_jv_j^*\right]$, 
and they allow one to write the nontrivial part of the grand canonical 
Hamiltonian in approximate diagonal form as 
$\hat H -\mu\hat N \sim \sum_j E_j \hat\alpha_j^{\dagger}\hat\alpha_j$. 
In the absence of the perturbation, Eq. \eqref{eq: BdG} is easily solved 
using complex exponentials parameterized by momentum $k$, yielding the 
unperturbed energies $E_k^{(0)}=\omega(k)$. The perturbation 
\eqref{eq: pert} separates modes of definite parity, so we write the 
unperturbed modes in a basis of cosine and sine functions: 
$u_{k,c}^{(0)}(x)=u_k\cos(kx)/\sqrt{L}$, $u_{k,s}^{(0)}(x)=u_k\sin(kx)/\sqrt{L}$, $v_{k,c}^{(0)}(x)=v_k\cos(kx)/\sqrt{L}$, 
and $v_{k,s}^{(0)}(x)=v_k\sin(kx)/\sqrt{L}$. The amplitudes $u_k$ and 
$v_k$ are defined by 
$u_k=~\sqrt{\frac{k^2/2+n_{1D}V(k)}{2\omega(k)}+\frac{1}{2}}$ 
and $v_k =~-\mathrm{sgn}(V(k))\sqrt{\frac{k^2/2+n_{1D}V(k)}{2\omega(k)}-\frac{1}{2}}$, for $k>0$ \cite{Fetter}. 
A system destabilizes when at least one of the excitation energies 
$E_j$ vanishes. 

Assuming an initially stable condensate, the influence of $U(x)$ on the excited states can also be approximated
in perturbation theory.
A perturbation theory for the BdG equations was developed rigorously 
in Refs. \cite{Gaul2011,Liang2008,Lugan2011} in the phase-density representation. Since we are chiefly concerned with first-order mode destabilization, 
we develop a tractable perturbation theory that paints a 
physical picture of mode softening and naturally encompasses the degeneracy 
of the roton spectrum. After expanding $E_k$, $u_k(x)$, and $v_k(x)$ 
in perturbation series, and then substituting into Eq \eqref{eq: BdG}, 
we derive the first-order perturbation equation for the corrections 
$E_k^{(1)}$, $u_k^{(1)}(x)$, and $v_k^{(1)}(x)$:
\begin{equation} \label{eq: BdG pert}
\begin{aligned}
&\begin{pmatrix}
\omega(k)+\frac{1}{2}\partial_x^2-X^{(0)} & -X^{(0)} \\
-X^{(0)} & -\omega(k)+\frac{1}{2}\partial_x^2-X^{(0)}
\end{pmatrix}
\begin{pmatrix}
u_k^{(1)} \\
v_k^{(1)}
\end{pmatrix} = \\&
\begin{pmatrix}
U_c^{(1)}+X^{(1)} & X^{(1)} \\
X^{(1)} & U_c^{(1)}+X^{(1)}
\end{pmatrix}
\begin{pmatrix}
u_k^{(0)}\\
v_k^{(0)}
\end{pmatrix}
-E_k^{(1)}
\begin{pmatrix}
u_k^{(0)} \\
-v_k^{(0)}
\end{pmatrix}
\end{aligned}
\end{equation}
where $X^{(0)}[\chi]=N\int dx'\chi(x')\psi^{(0)2}V(x-x')$ is the zeroth-order 
exchange integral operator, $U_c^{(1)}(x)$ is the first-order combined 
potential given by Eq. \eqref{eq: Uc}, and 
$X^{(1)}[\chi]=N\int dx'\chi(x')\psi^{(0)}V(x-x')\left(\psi^{(1)}(x')+\psi^{(1)}(x)\right)$ 
is the first-order exchange integral operator. We can isolate the energy 
shift $E_k^{(1)}$ on the right-hand side by taking advantage of the fact 
that the left-hand side vanishes whenever it is acted on by the 
operator $\int dx \left(u_{k'}^{(0)},v_{k'}^{(0)}\right)$ for any 
$\omega(k')=\omega(k)$. Degenerate perturbation theory is, in situations exhibiting a roton-maxon excitation spectrum, generally necessary for a complete understanding of first-order effects; however, we will see that the stability structure at $\lambda_{\mathrm{L}}=\lambda_{\mathrm{rot}}/2$ may be explained without accounting for degeneracy.
To simplify our notation, we denote the hermitian matrix on the right-hand 
side of \eqref{eq: BdG pert} by $\mathcal{A}$.

It turns out that the matrix elements of $\mathcal{A}$ vanish in most 
instances, simplifying our analysis. Firstly, the sine modes completely 
decouple from the cosine modes as anticipated. Moreover, the matrix 
element of $\mathcal{A}$ between any modes 
$\left(u_{k'}^{(0)},v_{k'}^{(0)}\right)$ and $\left(u_{k}^{(0)},v_{k}^{(0)}\right)^T$ vanishes unless the mode-matching condition $\left|k'\pm k  \right|=k_{\mathrm{L}}$ 
is satisfied.  This follows from reasons of orthogonality, since all matrix elements are evaluated by integrating products of three sine or cosine functions. Satisfying both the mode-matching condition and the degeneracy condition $\omega(k^{\prime})=\omega(k)$ severely limits the number of nonzero matrix elements determining the modes (or degenerate mode mixtures) that shift at first order.

\begin{figure}
\includegraphics[width = .85\columnwidth]{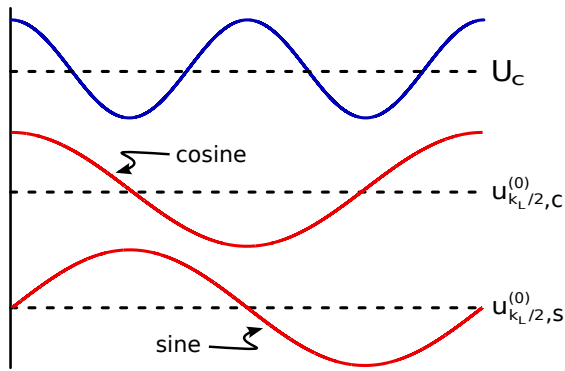}
\caption{
Depiction of the overlap between the combined potential $U_c(x)$ and the cosine and sine Bogoliubov modes for $k=k_L/2$.
} \label{fig: modes}
\end{figure}

The significance of the mode matching is illustrated schematically in Fig. \ref{fig: modes}. For any given perturbation probe with wave vector $k_{\mathrm{L}}$, a standing matter wave of this wavelength is established, defining the combined potential $U_c(x)$ (upper panel). Against this backdrop are shown the cosine and sine modes of wave vector $k_{\mathrm{L}}/2$ (during some phase of their oscillation) in the lower two panels. We depict these modes because they always satisfy both the mode-matching and degeneracy conditions. The density oscillations of the cosine mode accumulate atoms at the maxima of $U_c(x)$, causing an increase in the mode energy by an amount $E_{k_{\mathrm{L}}/2,c}^{(1)}=\mathcal{A}_{k_{\mathrm{L}}/2,k_{\mathrm{L}}/2}^{(c)}$. Conversely, the density oscillations of the sine mode collect atoms at the minima of $U_c(x)$, thereby lowering the mode energy by the amount $E_{k_{\mathrm{L}}/2,s}^{(1)}=-E_{k_{\mathrm{L}}/2,c}^{(1)}$. Linear combinations of degenerate modes satisfying $\left|k'\pm k\right|=k_{\mathrm{L}}$ may also accomplish this spatial appropriation of atoms in an {\it average} sense, as the direct terms in the diagonal elements appear as $\mathcal{A}_{j,j}\sim \int dx \left(u_j^{(0)2}(x)+v_j^{(0)2}(x)\right)U_c(x)$, where $u_j^{(0)}$ and $v_j^{(0)}$ are linear combinations of matched modes. This is the mechanism behind the first-order softening of Bogoliubov modes and degenerate mode mixtures, and it is similar to the discussion of ``staggered modes'' found in Ref. \cite{Morsch2006}.

Given that the roton mode is a local minimum of the system dispersion, we expect to observe a lower stability boundary for all lattice periodicities in which the roton softens to first order. This occurs when the roton mode-matches to itself ($k_{\mathrm{L}}=2k_{\mathrm{rot}}$) and when it mode-matches to a degenerate phonon ($k_{\mathrm{L}}\approx k_{\mathrm{rot}}$). These two scenarios correspond to the stability dips observed at $\lambda_{\mathrm{L}}=\lambda_{\mathrm{rot}}/2$ and $\lambda_{\mathrm{L}}\approx \lambda_{\mathrm{rot}}$, respectively. Of course, the latter structure was already expected due to strongly-enhanced, static density modulation, as discussed previously. The structure at $\lambda_{\mathrm{L}} = \lambda_{\mathrm{rot}}/2$, however, occurs in spite of relatively weak density modulation. In that case, the roton responds directly to the perturbation, rather than to a strongly-amplified mean field potential. 

We note that our discussions of density response and mode softening remain applicable despite the increase in quantum and thermal fluctuations for rotonized systems. There exists a minimum roton energy $\Delta$ below which either thermal or quantum fluctuations cause a strong reduction in the condensate fraction \cite{Boudjemaa2012}. For unperturbed dBECs with roton energies exceeding $\Delta$, the perturbing lattice will likely drive the system towards an uncondensed or thermal state before reaching dynamic instability. The criterion for the existence of a stable dBEC would then change from $min\{\omega(k)\}>0$ to $min\{\omega(k)\}>\Delta$, effectively lowering the boundaries in Fig. \ref{fig: stability} without changing the essential structure thereof.

In summary, we have explored the novel stability features of quasi-one-dimensional dipolar condensates in a weak lattice. Rotonized systems destabilize rapidly when the lattice periodicity $\lambda_{\mathrm{L}}$ matches either half of the roton wavelength or a subharmonic of the roton. The observation of these stability dips is a clear signal of rotonization, as well as a direct measurement the roton wavelength.

We thank B. Lev for fruitful discussions. J.P.C. acknowledges support 
from the US DoD through the NDSEG fellowship program. 
R.M.W. acknowledges support from an NRC postdoctoral fellowship. 
J.L.B acknowledges financial support from the NSF.

\bibliographystyle{amsplain}

\end{document}